\documentclass[namedreferences]{SolarPhysics}
\usepackage{epsfig}

\begin{document}
\begin{article}
\begin{opening}

\title{THE CITATION IMPACT OF DIGITAL PREPRINT\\ 
       ARCHIVES FOR SOLAR PHYSICS PAPERS\footnote{{\it Editors' Note}:  
This paper lies outside the normal purview of {\it Solar Physics} papers, 
however the editors feel that the content is of sufficient importance for 
all {\it Solar Physics} authors and readers to merit its publication.}}

\author{TRAVIS S. \surname{METCALFE}}

\runningauthor{METCALFE}
\runningtitle{CITATION IMPACT OF PREPRINT ARCHIVES}

\institute{High Altitude Observatory and Scientific Computing Division, 
NCAR,\\PO Box 3000, Boulder, CO 80307 U.S.A.\email{travis@ucar.edu}\\}

\date{Received 05 July 2006; accepted 11 October 2006}

\begin{abstract}
Papers that are posted to a digital preprint archive are typically cited 
twice as often as papers that are not posted. This has been demonstrated 
for papers published in a wide variety of journals, and in many different 
subfields of astronomy. Most astronomers now use the {\it arXiv.org} 
server ({\it astro-ph}) to distribute preprints, but the solar physics 
community has an independent archive hosted at Montana State University. 
For several samples of solar physics papers published in 2003, I quantify 
the boost in citation rates for preprints posted to each of these servers. 
I show that papers on the MSU archive typically have citation rates 1.7 
times higher than the average of similar papers that are not posted as 
preprints, while those posted to {\it astro-ph} get 2.6 times the average. 
A comparable boost is found for papers published in conference 
proceedings, suggesting that the higher citation rates are not the result 
of self-selection of above-average papers.
\end{abstract} 
\keywords{sociology of astronomy}

\end{opening}


\section{Background}

The {\it arXiv.org} preprint server ({\it astro-ph}\footnote{\tt 
http://arXiv.org/archive/astro-ph}) has been operating since 1992 
(Ginsparg, 2001), and now includes about 80\% of all new papers published 
in major refereed astrophysics journals around the globe (Metcalfe, 2005). 
Because it is the {\it single source} that many astronomers now use to 
keep up with the literature, papers that are posted to {\it astro-ph} are 
cited roughly twice as often as papers that are not posted. This has been 
shown to have little to do with the significance of the paper\,--\,even 
conference proceedings are cited twice as often when posted, though still 
20 times less overall (Schwarz and Kennicutt, 2004).

Falling somewhere between earth sciences and astrophysics, the solar 
physics community has been slower to adopt preprint archives than some 
subfields, with only about 7\% of {\it Solar Physics} papers posted to 
{\it astro-ph} in 2005, compared to about 79\% of papers published in {\it 
Astrophys.~J.}, and less than 1\% of papers published in {\it 
Geophys.~Res.~Lett.} An independent solar physics archive was established 
in 1999 at Montana State University (MSU\footnote{\tt 
http://solar.physics.montana.edu/cgi-bin/eprint/index.pl}), which has 
attracted a slightly larger following with 17\% of {\it Solar Physics} 
papers posted last year. The adoption rate is slightly higher among solar 
physicists who publish in broader astrophysics journals, with about 19\% 
of {\it Astron.~Astrophys.} papers and 29\% of {\it Astrophys.~J.} papers 
in this subfield posted to {\it astro-ph} in 2005.

With so much potential for growth in the adoption rate of preprint 
archives among the solar physics community, several questions arise: 
(1)~Do papers in this subfield enjoy a similar boost in citation rates 
when posted to one of these archives? (2)~Is there evidence that the boost 
is not due to self-selection of better papers, and (3)~Is there any 
quantitative reason to prefer one of the archives over the other? This 
paper aims to answer these questions with citation statistics obtained 
using NASA's Astrophysics Data System (Kurtz {\it et al.}, 2000) for three 
samples of papers published in 2003. This is far enough in the past to 
generate significant citation counts, while recent enough to sample a 
higher adoption rate for the preprint archives.


\section{Methodology and Results}

Although the citation database of the Astrophysics Data System (ADS) is 
not complete, it includes data from all of the major astronomy and physics 
journals beginning in 1999. Any incompleteness should have a uniform 
effect on all of the papers considered in this section.

\subsection{Can preprint archives boost citation rates?}

Our first sample includes the 171 papers published in {\it Solar Physics} 
during 2003, which have collectively received 512 citations that are 
included in the ADS database (as of 2006 June 15). To isolate the citation 
impact of the individual preprint archives we exclude one paper that was 
posted to both, which received a total of six citations. In 2003, just 
seven {\it Solar Physics} papers were posted exclusively to the MSU 
archive (receiving 33 citations), while six were posted only to {\it 
astro-ph} (43 citations)\,--\,leaving 157 unposted papers to account for 
the remaining 430 citations. Despite the small numbers, there is a 
significant difference in the average citation rates of these three sets 
of papers (see Table~I, where errors are assigned from $\sqrt{N}$ 
uncertainties on the citation counts). Compared to the sample of unposted 
papers, preprints posted to the MSU archive received an average of 
1.7$\pm$0.3 times as many citations while those posted to {\it astro-ph} 
received an average of 2.6$\pm$0.4 times as many. Schwarz and Kennicutt 
(2004, their Figure~9) found no significant difference in the long-term 
citation patterns of papers posted before and after peer review, so this 
citation boost is not merely due to the slightly earlier availability of 
the paper.

\begin{table*}[t]
\begin{center}
\caption{The citation impact of digital preprint archives for solar 
physics papers.}
\begin{tabular}{lcrccrccrc}
\hline
Sample&&\multicolumn{2}{c}{unposted}&&\multicolumn{2}{c}{MSU~archive}&&
\multicolumn{2}{c}{\it astro-ph}\\
&&\#&$\langle$citations$\rangle$&&\#&$\langle$citations$\rangle$&&
\#&$\langle$citations$\rangle$\\
\hline
{\it Solar~Physics}\dotfill && 157 & $2.7\pm0.1$   && 7 & $4.7\pm0.8$ && 
6 & $7.2\pm1.1$    \\
IAU~Symp.~210$\ldots$ && 150 & $0.24\pm0.04$ && 0 & $\cdots$    &&
20 & $0.65\pm0.18$ \\
MSU~archive\dotfill   && 0 & $\cdots$ && 82 & $7.6\pm0.3$       &&
20 & $10.2\pm0.7$  \\
\hline
\end{tabular}
\end{center}
\end{table*}

\subsection{Are the citation rates higher from self-selection?}

Confronted with the data in Table~I, many scientists will wonder whether 
there is a self-selection effect. Perhaps the papers on the preprint 
archives are not representative of the full sample, and the authors chose 
to post them because they were better than average? This could explain why 
these papers received more citations.

We can address this question by examining a second sample of papers that 
we would not expect authors to self-select for high quality. Schwarz and 
Kennicutt (2004) found that conference proceedings are typically cited 
about 20 times less than refereed journal papers, so we choose a solar 
physics proceedings published in 2003, from IAU Symposium 210 {\it 
``Modelling of Stellar Atmospheres''}. Of the 170 papers included in this 
volume, only 20 were posted to {\it astro-ph}, and none were posted to the 
MSU archive. As expected, the overall rate of citations is much 
lower\,--\,with 13 citations to the 20 posted papers, and 36 citations to 
the 150 unposted papers (see Table~I). Even conference proceedings papers, 
though cited an order of magnitude less overall, are cited 2.7$\pm$0.9 
times more when posted to {\it astro-ph}. This is consistent with the 
factor of 2.6$\pm$0.4 we found for {\it Solar Physics} papers.

\subsection{Which preprint archive should solar physicists use?}

It is clear that both preprint archives provide a significant boost in the 
citation rates of posted papers, but the evidence suggests that {\it 
astro-ph} provides a slightly larger boost than the MSU archive. We can 
perform an additional test of the relative impact of the two archives by 
looking at a third sample of papers. There were 102 refereed papers posted 
to the MSU archive in 2003, and 20 were also posted to {\it astro-ph}. The 
papers posted to {\it both} preprint archives received 204 citations, 
while the 82 papers posted only to the MSU archive garnered 621 citations 
(see Table~I). This implies an additional boost of 1.3$\pm$0.1 for papers 
posted to {\it astro-ph}, on top of the factor of approximately two boost 
from posting to the MSU archive. This is roughly consistent with the 
citation impact of {\it astro-ph} alone.

Note that the average citation rate for {\it all} refereed papers on the 
MSU archive (most of them published in broader astrophysics 
journals\footnote{Among papers in the MSU sample, 41\% were published in 
{\it Astrophys.~J.}, 25\% in {\it Astron.~Astrophys.}, and 11\% in {\it 
Solar Physics}. There was no significant difference in the adoption rate 
of {\it astro-ph} for these three sub-samples.}) is systematically higher 
than that for {\it Solar Physics} papers alone. In fact, astrophysics 
papers posted to the MSU archive have citation rates comparable to {\it 
Solar Physics} papers posted to {\it astro-ph}. This implies that the 
higher citation rates are connected to awareness of the paper among the 
broader astrophysics community\,--\,either by publication in an 
astrophysics journal, or through a preprint posted to {\it astro-ph}. The 
highest average citation rates occur for papers that have been posted to 
{\it both} preprint archives, though {\it astro-ph} appears to be the best 
{\it single} choice.


\section{Conclusions and Discussion}

Despite the slower adoption by the solar physics community, digital 
preprint archives boost the citation rates of posted papers to twice the 
level of unposted papers, a conclusion first noted in the comprehensive 
study of Schwarz and Kennicutt (2004). The evidence suggests that, like 
many other subfields in astronomy, the citation rate is elevated from the 
improved visibility of the paper rather than from self-selection by 
authors choosing to post above-average papers. Unlike other subfields, 
solar physicists maintain an independent preprint archive which also 
boosts citation rates, though the broader user-base of {\it astro-ph} 
provides a larger boost.

If citation rates track the assimilation of new results by the community, 
then {\it astro-ph} seems to be the best single form of communication 
available. Editors who want to maximize the impact of their journals 
should encourage authors to post their preprints to {\it astro-ph}. 
Authors in solar physics, where {\it astro-ph} is currently underutilized, 
should consider the advantages that other subfields have already 
discovered.


\acknowledgements
I would like to thank Sarah Gibson for a comment that helped motivate this 
work, and Stan Solomon for several constructive suggestions. This research 
made use of NASA's Astrophysics Data System. The National Center for 
Atmospheric Research is a federally funded research and development center 
sponsored by the National Science Foundation.


\end{article} 

\end{document}